\documentclass[reprint,preprintnumbers,longbibliography]{revtex4-2}
\usepackage{textgreek}
\usepackage{braket}
\usepackage{physics}
\usepackage{multirow}
\usepackage{siunitx}

\usepackage{tikz}
\usetikzlibrary{decorations.markings, arrows, positioning, calc}

\usepackage{graphicx,times}
\usepackage{latexsym}
\usepackage{mathtools}

\usepackage{amsmath,amssymb,amsbsy,amsfonts}
\usepackage{array}
\usepackage{bm}

\usepackage{graphics}
\graphicspath{{./fig/}}

\usepackage{mathrsfs}
\usepackage{xcolor}
\usepackage{cancel}
\usepackage[normalem]{ulem}

\usepackage[unicode, pdfprintscaling=None, colorlinks]{hyperref}
\hypersetup{
    colorlinks,
    allcolors=blue!50!black,
}
\usepackage{inconsolata}

\usepackage[capitalise]{cleveref}

\renewcommand\>{\rangle}
\newcommand\<{\langle}

\usepackage{relsize}

\usepackage{microtype}

\usepackage{xfrac}
\usepackage{orcidlink}

\usepackage{ulem} 
\usepackage{amssymb}

\newcommand\del\partial
\usepackage{relsize}

\usepackage[acronyms, smallcaps, nohypertypes={acronym}, nopostdot, style=super, nonumberlist, toc]{glossaries}
\glsenableentrycount
\makeglossaries  
\setacronymstyle{long-sc-short}

\newcommand\ac[1]{\gls{#1}}

\newcommand\acp[1]{\glspl{#1}}

\newacronym{WF}{wf}{Wilson-Fisher}
\newacronym{AF}{af}{asymptotically free}
\newacronym{RG}{rg}{renormalization group}
\newacronym{WZW}{wzw}{Wess-Zumino-Witten}
\newacronym[longplural={conformal field theories}]{CFT}{cft}{conformal field theory}
\newacronym[longplural={lattice field theories}]{LFT}{lft}{lattice field theory}
\newacronym[longplural={effective field theories}]{EFT}{eft}{effective field theory}
\newacronym[longplural={quantum field theories}]{QFT}{qft}{quantum field theory}
\newacronym{LEC}{lec}{low-energy constant}
\newacronym{QCD}{qcd}{quantum chromodynamics}
\newacronym{MC}{mc}{Monte Carlo}
\newacronym{IR}{ir}{infrared}
\newacronym{UV}{uv}{ultraviolet}
\newacronym{SNR}{snr}{signal-to-noise ratio}
\newacronym{NLSM}{nl$\sigma$m}{nonlinear sigma model}
\newacronym{CSA}{csa}{Cartan subalgebra}
\newacronym{SSB}{ssb}{spontaneous symmetry breaking}
\newacronym{DOF}{dof}{degrees of freedom}
\newacronym{DMRG}{dmrg}{density matrix renormalization group}
\newacronym{PBC}{pbc}{periodic boundary conditions}
\newacronym{OBC}{obc}{open boundary conditions}
\newacronym{TDVP}{tdvp}{Time-Dependent Variational Principle}

\begin{document}
\title{Preparation for Quantum Simulation of the 1+1D O(3) Non-linear $\sigma$-Model using Cold Atoms}

\author{Anthony N. Ciavarella \,\orcidlink{0000-0003-3918-4110}}
\email{aciavare@uw.edu}
\author{Stephan Caspar\,\orcidlink{0000-0002-3658-9158}}
\email{caspar@uw.edu}
\author{Hersh Singh\,\orcidlink{0000-0002-2002-6959}}
\email{hershsg@uw.edu}
\author{Martin J. Savage\,\orcidlink{0000-0001-6502-7106}}
\email{mjs5@uw.edu}
\affiliation{InQubator for Quantum Simulation (IQuS), Department of Physics, University of Washington, Seattle, Washington 98195-1550, USA}
\preprint{IQuS@UW-21-038,  INT-PUB-22-051}

\begin{abstract}
The 1+1D O(3) non-linear $\sigma$-model is a model system for future quantum lattice simulations of other asymptotically-free theories, such as non-Abelian gauge theories. 
We find that utilizing dimensional reduction can make efficient use of  two-dimensional layouts presently available on cold atom quantum simulators. 
A new definition of the renormalized coupling is introduced, which is 
applicable to systems with open boundary conditions 
and can be measured using analog quantum simulators.
Monte Carlo and tensor network calculations are performed to determine the quantum resources required to reproduce perturbative short-distance observables. 
In particular, we show that a rectangular array of 48 Rydberg atoms with existing quantum hardware capabilities should be able to adiabatically prepare low-energy states of the perturbatively-matched theory. 
These states can then be used to simulate non-perturbative observables in the 
continuum limit that lie beyond the reach of classical computers.
\end{abstract}

\maketitle
\glsresetall

\section{Introduction}
Future quantum simulations of Abelian and non-Abelian \acp{QFT}, 
such as \ac{QCD}, 
and descendant effective field theories,
will be important in developing robust predictive capabilities of the 
dynamics in a variety of physical systems of importance in high-energy and nuclear physics,
ranging from the early universe,
to highly-inelastic processes in particle colliders,
to the evolution of extreme astrophysical environments.
Beyond the capabilities of classical computation, these challenges can only be addressed using yet-to-be-engineered quantum computers of sufficient capability~\cite{Feynman1982,Benioff:1980}.
During the last decade, rapid advances in the control of 
coherence and entanglement in the laboratory
has led to the deployment of the first generation of quantum computing platforms, 
built around superconducting qubits~\cite{Huang_2020s,PhysRevLett.105.223601,wang2022high,wallraff2004strong,chiorescu2004coherent}, trapped ions~\cite{2019Trap}, and neutral atoms~\cite{Henriet_2020,Browaeys_2020,Barredo_2020,bluvstein2022quantum}. 
These can be operated in a digital manner, where a universal gate-set is used to emulate a specific Hamiltonian, or an analog manner where the system is tuned to 
natively implement a target Hamiltonian, or as quantum annealers~\cite{johnson2011quantum,PhysRevB.82.024511,6802426}. 
While digital quantum simulation platforms are universal in the sense that they can simulate an arbitrary Hamiltonian, 
the difficulties of implementing quantum gates has so far 
limited digital quantum simulations to relatively small systems. 
In contrast, analog quantum simulations have been performed with larger systems, 
but are limited by the native Hamiltonian of the experimental platform. 
Recent work has indicated that error rates on some analog simulation platforms 
are low enough for potential quantum advantages in physically interesting systems 
to be within reach~\cite{flannigan2022propagation}.
In particular, 
cold atom systems have  been used to simulate the dynamics of quantum systems 
in regimes that are difficult for classical computers to simulate~\cite{ebadi2021quantum,Scholl_2021}.

With the emerging potential of quantum computers, 
and the known limitation of classical computing, 
a growing effort is underway to develop efficient 
mappings of \acp{QFT} onto quantum computers, 
and the time-evolution of an array of initial conditions.
The asymptotic freedom of SU(2) and SU(3) gauge theories enables spatial lattice calculations to be perturbatively close to the continuum, and systematically correctable,
as has long be used for lattice \ac{QCD} classical simulations.
Traditional lattice mappings of gauge theories, 
such as Kogut-Susskind~\cite{PhysRevD.11.395},
have led to first calculations of modest systems in low-dimensions 
in U(1)~\cite{Martinez:2016yna,Klco:2018kyo,Lu:2018pjk,Surace_2020,Nguyen:2021hyk,Thompson:2021eze,PhysRevLett.122.050403}, SU(2)~\cite{Klco:2019evd,ARahman:2021ktn,Atas:2021ext,Rahman:2022rlg} and SU(3)~\cite{Ciavarella:2021nmj,Ciavarella:2021lel,Farrell:2022wyt,Farrell:2022vyh,Atas:2022dqm}, and estimates of resource requirements,
along with improved understandings about how to move forward.
These advances have also driven the development of new and different 
encodings of \acp{QFT} onto finite discrete degrees of freedom~\cite{Brower:1997ha,PhysRevA.73.022328,Zohar:2011cw,Zohar:2012ay,Tagliacozzo:2012vg,Banerjee:2012xg,Tagliacozzo:2012df,Zohar:2012xf,Zohar:2012ts,Wiese:2013uua,Hauke:2013jga,Marcos:2014lda,Kuno:2014npa,Bazavov:2015kka,Kasper:2015cca,Brennen:2015pgn,Kuno:2016xbf,Martinez:2016yna,Zohar:2016iic,Kasper:2016mzj,Muschik:2016tws,Gonzalez-Cuadra:2017lvz,Banuls:2017ena,Dumitrescu:2018njn,Klco:2018kyo,Kaplan:2018vnj,Kokail:2018eiw,Lu:2018pjk,Yeter-Aydeniz:2018mix,Stryker:2018efp,Gustafson:2019mpk,cloet2019opportunities,Bauer:2019qxa,Shehab:2019gfn,Klco:2019xro,Alexandru:2019nsa,Davoudi:2019bhy,Avkhadiev:2019niu,Klco:2019evd,Magnifico:2019kyj,Gustafson:2021mky,Banuls:2019bmf,Klco:2019yrb,Mishra:2019xbh,Luo:2019vmi,Kharzeev:2020kgc,Mueller:2020vha,Shaw2020quantumalgorithms,PhysRevLett.122.050403,Yang_2020,Ji:2020kjk,Bender:2020jgr,Haase:2020kaj,Halimeh:2020ecg,Robaina:2020aqh,Yeter-Aydeniz:2020jte,Paulson:2020zjd,Halimeh:2020djb,VanDamme:2020rur,Barata:2020jtq,Milsted:2020jmf,Kasper:2020owz,Ott:2020ycj,Ciavarella:2021nmj,Bauer:2021gup,Gustafson:2021imb,ARahman:2021ktn,Atas:2021ext,Yeter-Aydeniz:2021olz,Davoudi:2021ney,Kan:2021nyu,Stryker:2021asy,Aidelsburger:2021mia,Zohar:2021nyc,Halimeh:2021vzf,Yeter-Aydeniz:2021mol,Knaute:2021xna,Wiese:2021djl,Meurice:2021pvj,Mueller:2021gxd,Riechert:2021ink,Halimeh:2021lnv,Zhang:2021bjq,Alam:2021uuq,Deliyannis:2021che,Perlin:2021xux,Funcke:2021aps,Gustafson:2021jtq,VanDamme:2021njp,Thompson:2021eze,Kan:2021blb,Ashkenazi:2021ieg,Alexandrou:2021ynh,Bauer:2021gek,Wang:2021iox,Iannelli:2021jhs,Ciavarella:2021lel,Nguyen:2021hyk,Yeter-Aydeniz:2022vuy,Hartung:2022hoz,Illa:2022jqb,Ji:2022qvr,Carena:2022kpg,Halimeh:2022rwu,Mildenberger:2022jqr,Deliyannis:2022uyh,Ciavarella:2022zhe,Caspar:2022llo,Bauer:2022hpo,Halimeh:2022pkw,Halimeh:2022mct,Raychowdhury:2022wbi,Dreher:2022scr,Rahman:2022rlg,Greenberg:2022kzy,Tuysuz:2022knj,Farrell:2022wyt,Atas:2022dqm,Bringewatt:2022zgq,Grabowska:2022uos,Asaduzzaman:2022bpi,Carena:2022hpz,Gustafson:2022xdt,Davoudi:2022uzo,Avkhadiev:2022ttx,Jang:2022nun,Farrell:2022vyh,https://doi.org/10.48550/arxiv.2206.12454}. 

Interestingly, the O(3) \ac{NLSM} in $1+1$ dimensions 
is a theory of interacting scalar particles that is asymptotically free, 
and can support a topologically non-trivial ground state (vacuum).
Because of these qualitative similarities with \ac{QCD}, 
it serves as a useful test-bed for the 
development of computational methods for \ac{QCD}. 
A number of mappings of the O(3) \ac{NLSM} suitable for quantum simulation have been introduced, 
including the Heisenberg comb, fuzzy sphere, 
angular momentum truncations and D-theory~\cite{https://doi.org/10.48550/arxiv.1911.12353,https://doi.org/10.48550/arxiv.2210.03679,Bhattacharya:2020gpm,CHANDRASEKHARAN2002388,Brower:2003vy,Caspar:2022llo}. 
Previous work has shown that at lowest truncation, the fuzzy sphere regularization reproduces the O(3) \ac{NLSM}~\cite{https://doi.org/10.48550/arxiv.2209.00098}, while the angular momentum truncation requires a larger local Hilbert space to do so~\cite{PhysRevD.99.074501}. 
The D-theory mapping with \ac{PBC} has been shown in a number of works to reproduce the 
O(3) \ac{NLSM}, both with and without a $\theta$-term~\cite{CHANDRASEKHARAN2002388,Brower:2003vy,Caspar:2022llo}.
However, present-day analog simulators, including 
arrays of cold atoms, only support \ac{OBC}.

A central ingredient in lattice 
simulations of asymptotically-free \acp{QFT} 
is the perturbative matching between the continuum  and the lattice at short-distances (compared to the scale at the theory becomes non-perturbative).
In this work, 
it is shown that it is possible to perform this matching for the O(3) \ac{NLSM} on existing analog quantum simulators. 
A definition of the renormalized coupling 
in the O(3) \ac{NLSM} that is
suitable to be used with \ac{OBC} is introduced, and 
implemented using tensor network simulations 
to compute the step-scaling function in the D-theory mapping. 
The step-scaling function is then matched to perturbative results 
at short distances (ultraviolet), 
and the results of Monte Carlo calculations at long-distances (infrared),
allowing for the minimum number of qubits required to reproduce 
continuum physics of the O(3) \ac{NLSM} (to a given level of precision)
to be determined.
Tensor-network simulations  
indicate 
that asymptotic freedom and  
non-perturbative dynamics beyond the capabilities of classical computers
in the O(3) \ac{NLSM}
can be potentially simulated with current cold-atom experimental configurations.

%%%%%%%%%%%%%%%%%%%%%%%%%%%%%%%%%%%%%%%%%%
\section{Mapping D-Theory to Qubit Registers}
The $1+1$D O(3) \ac{NLSM} is defined by the action
\begin{equation}
    S = \frac{1}{2 g} \int\!\! dt\ dx \ \partial_\mu \Vec{\phi}(x,t) \cdot \partial^\mu \Vec{\phi}(x,t) \ \ \ , 
    \label{eq:O3action}
\end{equation}
where $\Vec{\phi}(x,t)$ is a vector of three scalar fields subject to the constraint $\Vec{\phi}(x,t) \cdot \Vec{\phi}(x,t) = 1$. 
This constraint is responsible for transforming the free-boson action in 
Eq.~(\ref{eq:O3action}) into an interacting asymptotically-free \ac{QFT}. 

This theory has been extensively studied using classical \ac{MC} methods using a straightforward discretization of the above continuum action,
\begin{equation}
  S_{\text{lat}} = -\frac{1}{g}\sum_{\< i j \>} \Vec{\phi}_{i} \cdot \Vec{\phi}_{j}  \ \ \ .
\end{equation} 
where the sum is over all nearest-neighbor sites $i,j$ on a square Euclidean spacetime lattice.

Simulating this theory on a quantum computer requires a truncation of the field,
and the D-theory formulation provides a natural mapping onto qubit degrees of freedom,
and an intrinsic truncation, utilizing dimensional reduction.
In this mapping, 
spin-$\frac{1}{2}$ degrees of freedom are placed on a 2D rectangular lattice of length $L_x$ sites in the $x$ direction and $L_y$ sites in the $y$ direction and coupled through an antiferromagnetic Heisenberg interaction, i.e.,
\begin{equation}
    \hat{H}^D = J_x \sum_{x,y} \Vec{S}_{x,y} \cdot \Vec{S}_{x+1,y} + J_y \sum_{x,y} \Vec{S}_{x,y} \cdot \Vec{S}_{x,y+1} \ \ \ .
    \label{eq:DTheoryNN}
\end{equation}
To obtain the 1+1D O(3) \ac{NLSM}, we choose $J_x$, $J_y$ such that the 2D model is in a massless (symmetry broken) phase when $L_x, L_y \to \infty$. With these choice of parameters, the continuum limit of the \ac{NLSM} is obtained in the limit  $L_x \gg L_y \gg 1$, as has been demonstrated in several previous works for $J_x=J_y$ \cite{PhysRevB.39.2344,Brower:2003vy,beard_efficient_2006,CHANDRASEKHARAN2002388,Caspar:2022llo}. This has enabled classical Monte Carlo studies of the O(3) \ac{NLSM} at finite density~\cite{PhysRevB.39.2344} and with a $\theta$ term \cite{Caspar:2022llo} without a sign problem. 
%
%Previous work has shown that when $J_x=J_y$ and when the lattice is 
% highly asymmetric $L_x \gg L_y$, dimensional reduction occurs, 
%and the low-energy degrees of freedom of this theory are described by the $1+1$D O(3) \ac{NLSM}~\cite{}. 
In the isotropic ($J_x=J_y$) D-theory approach, 
each even $L_y$ corresponds to a fixed coupling,
and as the correlation length grows exponentially in $L_y$, 
this corresponds to a coarse set of lattice spacings. 
A more refined set of lattice spacings can be explored by varying $J_x/J_y$. 
In the regime $J_x/J_y \lesssim 1$, dimensional reduction should still occur, 
while the correlation length is reduced.  

% due to the smaller value of $J_x/J_y$.

Determining the lattice spacing (in physical units) in any simulation
of a \ac{QFT} requires matching 
one or more dimensionful quantities calculated in lattice units to the corresponding experimentally or theoretically determined quantity.
Such determinations have associated systematic errors due to the finite volume, imprecise input parameters, and other effects, see for example  Ref.~\cite{Beane:2014oea}.
For the O(3) \ac{NLSM}, the renormalized coupling can be used to set the length scale. Typically, Monte Carlo studies of the O(3) \ac{NLSM} have been performed in a Euclidean spacetime with \ac{PBC}, and the renormalized coupling, $\Bar{g}(L)$, 
is defined in terms of two-point spacetime correlation functions projected onto momentum modes 
\cite{PhysRevLett.75.1891}.
This definition is somewhat problematic for 
our present purposes because quantum simulation platforms  do not  have direct access to Euclidean spacetime correlation functions, and, further, 
it is more natural to implement \ac{OBC} (for which momentum modes are no longer non-interacting eigenstates)
on current platforms.
Previous work has explored renormalized couplings defined in terms of energy gaps with \ac{OBC}~\cite{https://doi.org/10.48550/arxiv.2209.00098}. However, this is resource intensive to extract in practice on hardware, as it requires accurate preparation of both the ground state and first excited state and measurements of their energies.
In this work, we introduce a new definition of $\Bar{g}(L)$, given  in terms of spatial correlations, that recovers the traditional definition in the perturbative regime,
and which can be practically implemented in quantum simulations.
Explicitly, $\Bar{g}(L)$ is defined by
\begin{equation}
    \Bar{g}(L) = \frac{1}{2} \sqrt{\frac{1}{L\sin\left(\frac{\pi}{2L}\right)} \left(\frac{G_0}{G_1} - 1\right)} 
    \ \ \ ,
    \label{eq:CorrelationEigen}
\end{equation}
where $G_0$ and $G_1$ are the largest and second largest eigenvalues of the vacuum correlation matrix, $G_{x_1,x_2}$, defined by
\begin{equation}
    G_{x_1,x_2} = \sum_{y_1,y_2} (-1)^{x_1+y_1 + x_2 +y_2} \bra{\psi} \hat{S}^z_{x_1,y_1} \hat{S}^z_{x_2,y_2} \ket{\psi} 
    \  ,
\end{equation}
where $\ket{\psi}$ is the vacuum state of the Hamiltonian in Eq.~(\ref{eq:DTheoryNN}), 
and $\hat{S}^z_{x,y}$ is the $z$-component of the spin operator at site $(x,y)$.
Recently, another method to extract the running coupling on quantum platforms for 2+1D quantum electrodynamics was proposed in Ref.~\cite{https://doi.org/10.48550/arxiv.2206.12454}, albeit with \ac{PBC}.

To show that the continuum physics of the O(3) \ac{NLSM} can be recovered on a quantum device, we compute a  
universal step-scaling function, $F_s(z)$, defined as
\newcommand\gbare{g_{\text{bare}}}
\begin{equation}
    F_s(z) = s\frac{\bar g(s L, \gbare)}{\bar g(L, \gbare)} 
    \ \ \ ,
\end{equation}
where $z = \bar g(L, \gbare)$. Here, we emphasize that the bare coupling $\gbare$ is kept fixed on the right hand side.
In the limit $z\rightarrow0$, $F_s(z)$ probes \ac{IR} physics and in the $z\rightarrow \infty$ limit, $F_s(z)$ probes \ac{UV} physics. Therefore, if a lattice regularization reproduces the entire step scaling function it can be said to reproduce the continuum physics of the O(3) \ac{NLSM}. Any lattice regularization should be able to bridge the gap between perturbative \ac{UV} physics and the non-perturbative \ac{IR} physics. 
For simulations of asymptotically-free theories, 
it is essential to match the lattice theory to the continuum theory (\ac{UV})  
with as few computational resources as possible, 
as the resulting non-perturbative 
\ac{IR} physics emerges at parametrically larger length scales.

\begin{figure}
    \centering
    \includegraphics[width=8.6cm]{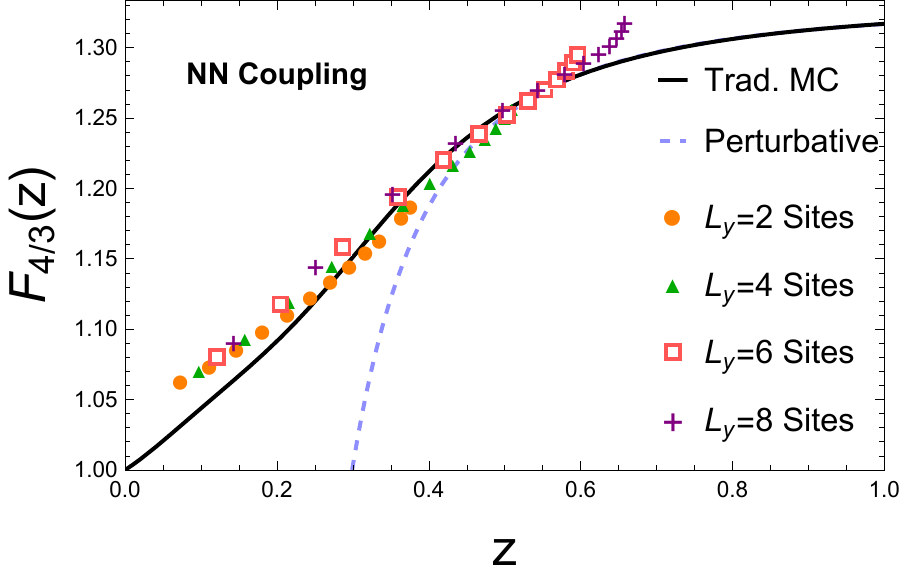}
    \caption{The step scaling function $F_\frac{4}{3}(z)$ for the coupling in Eq.~(\ref{eq:CorrelationEigen}) computed by varying $\frac{J_x}{J_y}$ for the nearest neighbor (NN) D-theory Hamiltonian for going from a lattice of size $6\times L_y$ sites to $8\times L_y$ sites. 
    The black line is a fit to results of Monte Carlo calculations 
    using the traditional lattice regularization.
    The dashed blue line is the perturbative result~\cite{PhysRevLett.75.1891}.
    }
    \label{fig:StepScaleNN}
\end{figure}
To determine the size of lattices required 
to reproduce the O(3) \ac{NLSM}, 
\ac{DMRG} calculations 
were performed using the C++ {\tt ITensor} library \cite{itensor-r0.3, itensor} to obtain the vacuum state of the Hamiltonian in Eq.~(\ref{eq:DTheoryNN}) 
for lattices of size $6\times L_y$ and $8\times L_y$ with \ac{OBC}~\cite{itensor,PhysRevLett.69.2863,PhysRevB.48.10345,2004MPS}. 
The renormalized couplings defined by Eq.~(\ref{eq:CorrelationEigen}) 
were used to compute $F_s(z)$ with $s=\frac{4}{3}$. 
Note that while traditionally $F_s(z)$ is computed for $s=2$, 
any value of $s$ may be used in principle,
and we have used $s=\frac{4}{3}$ to reduce the classical computing overhead.
Different points on the $F_\frac{4}{3}(z)$ curve,
shown in Fig.~\ref{fig:StepScaleNN},
were computed by varying 
$\frac{J_x}{J_y}$ in the range $0.1\leq \frac{J_x}{J_y} \leq 1.3$.
At the lower end of the perturbative regime, $z \lesssim  0.55$,
$F_s(z)$ is reproduced sufficiently well with $L_x =6,8$ lattice sites, 
provided a large transverse direction $L_y = 8$ is used.
This indicates that perturbative matching between the 
continuum and lattice O(3) \ac{NLSM} theories can be accomplished with as few as $64$ qubits on a quantum device.

\begin{figure}
    \centering
    \includegraphics[width=8.6cm]{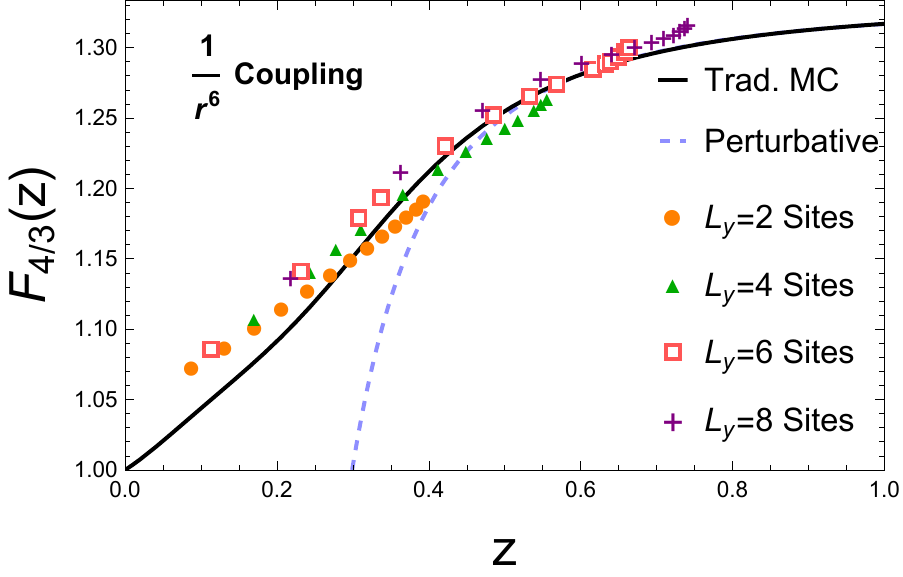}
    \caption{The step-scaling function computed by varying $\frac{a_x}{a_y}$ for the $\frac{1}{r^6}$ D-theory Hamiltonian for going from a lattice of size $6\times L_y$ sites to $8\times L_y$ sites.}
    \label{fig:StepScaleYR6}
\end{figure}
While the D-theory Hamiltonian with nearest-neighbor couplings is natural to consider, some quantum simulation platforms, such as cold atoms, have long range couplings. For example, arrays of Rydberg atoms with an s-wave coupling are described by a Hamiltonian with the form,
\begin{equation}
    \hat{H}^\text{Ryd.} 
    = \sum_i \frac{\Omega_i(t)}{2} \hat{X}_i 
    + \sum_i \Delta_i(t) \hat{n}_i 
    + \sum_{i<j} \frac{C_6 \ \hat{n}_i \hat{n}_j }{\abs{\Vec{x}_i 
    - \Vec{x}_j}^6} 
    \ \ \ ,
\end{equation}
where $\hat{n}_i$ is the Rydberg-state occupation of atom $i$, 
$\Vec{x}_i$ is the position of atom $i$, 
and $\hat{X}_i$ couples the ground state of atom $i$ to its excited Rydberg state~\cite{Henriet_2020,Browaeys_2020}. 
$\Omega_i(t)$ specifies the strength of the driving field at atom $i$, 
and $\Delta_i(t)$ specifies a local detuning. 
By identifying the excited-state occupation number with the $z$-component of a spin, 
it can be seen that this system is described by an 
Ising Hamiltonian with long-range interactions and time-dependent external fields. Due to this native encoding of the Ising model, 
Rydberg atoms have been used in a number of studies to perform 
analog quantum simulations of the Ising model~\cite{ebadi2021quantum,doi:10.1126/science.abi8794,bernien2017probing,Scholl_2021}. As we have shown in previous works, 
the Ising model with a strong transverse and longitudinal field can reproduce 
the dynamics of the Heisenberg model, and time dependent 
external fields can be used to adiabatically 
prepare ground states of the Heisenberg model with long range interactions~\cite{https://doi.org/10.48550/arxiv.2207.09438,https://doi.org/10.48550/arxiv.2210.04965}. In particular, by arranging atoms in a rectangular lattice and identifying the number operator of the atom at site $(x,y)$, 
$\hat{n}_{x,y}$ with a staggered $z$-component of a spin operator, 
i.e., $\hat{n}_{x,y} = \frac{1}{2} +(-1)^{x+y} \hat{S}^z_{x,y}$, 
it is possible to engineer a Heisenberg Hamiltonian,
\begin{align}
    \hat{H}^{\text{D6}} = \sum_{x_1,y_1,x_2,y_2} & \frac{(-1)^{1+x_1+y_1+x_2+y_2}}{\left( a_x^2(x_1-x_2)^2 + a_y^2(y_1-y_2)^2 \right)^3} \nonumber \\ 
    & \Vec{S}_{x_1,y_1} \cdot \Vec{S}_{x_2,y_2}  \ \ \ ,
    \label{eq:DTheoryR6}
\end{align}
where $a_{x,y}$ are the lattice spacings in the $x,y$ directions. The staggered identification of the number operator with the spin operator is necessary to ensure that the state with all atoms in their ground state, in which the system will begin in a quantum simulation, corresponds to a state with staggered spins that is adiabatically connected to the ground state of Eq.~\ref{eq:DTheoryR6}. The staggering identification also makes the long range interactions frustration-free. Note that the Hamiltonian implemented on hardware will differ from that of Eq.~\ref{eq:DTheoryR6} by a sign, but due to time reversal symmetry this does not present an issue. This Hamiltonian is equivalent to the Hamiltonian in Eq.~(\ref{eq:DTheoryNN}) with the addition of long-range frustration-free Heisenberg interactions. 
Therefore, it is expected that $a_{x,y}$ can be tuned so that dimensional reduction occurs and the low energy degrees of freedom are described by the $1+1$D O(3) \ac{NLSM}. 
To verify this, the step-scaling function for the vacuum state of this Hamiltonian was computed using \ac{DMRG}, with the results shown in Fig.~\ref{fig:StepScaleYR6},
where $\frac{a_y}{a_x}$ was varied in the range 
$0.1 \leq \left(\frac{a_y}{a_x}\right)^6 \leq 1.3$. 
The step-scaling function computed with $L_y=6$ 
reproduces the perturbative function over a range of parameters well 
into the perturbative regime,
demonstrating that, for this range of couplings, the \ac{UV} physics of the O(3) \ac{NLSM} is correctly reproduced. 
It is interesting to note that $L_y=6$ with nearest neighbor couplings only 
is not able to reproduce the step-scaling function as precisely in this region,
and in this sense, the $\frac{1}{r^6}$ coupling 
effectively implements an ``improved'' Hamiltonian
that enables more precise matching with fewer qubits. 
However, $L_y=6$ appears to be an optimum in this case, since $L_y=8$ has again larger systematic errors for this $L_x$.

\begin{figure}
    \centering
    \includegraphics[width=8.6cm]{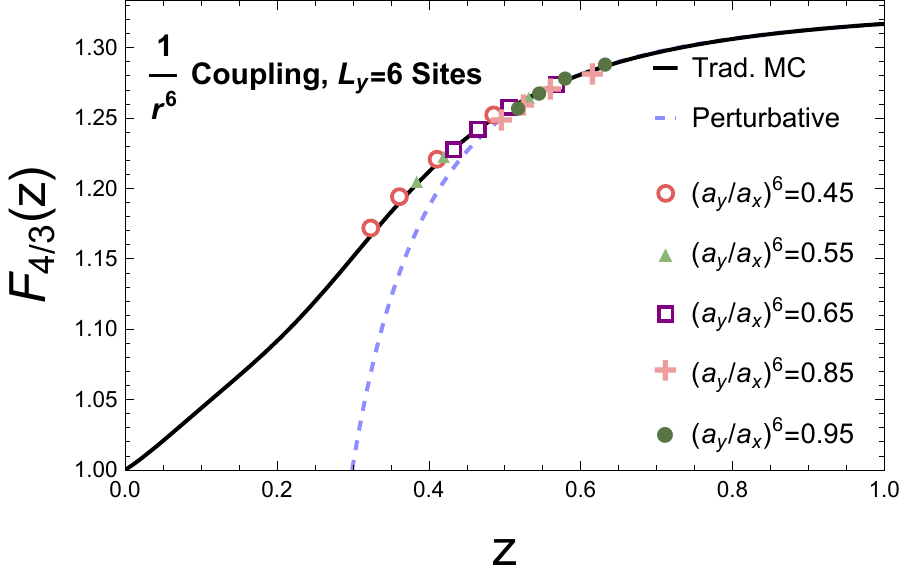}
    \caption{
    The step-scaling function computed for $L_x=6,12,18$ and $24$ sites 
    with the $\frac{1}{r^6}$ D-theory Hamiltonian with $L_y=6$ sites.
    }
    \label{fig:StepScaleLR6}
\end{figure}
With controlled matching to the continuum theory, non-perturbative \ac{IR} physics of the 
O(3) \ac{NLSM} 
is expected to be able to be simulated by keeping the Hamiltonian parameters $J_x,J_y,L_y$ fixed 
while increasing the lattice size $L_x$. 
To demonstrate that this procedure reproduces the \ac{IR} correctly, 
$F_s(z)$ was computed with \ac{DMRG} for lattices with 
larger $L_x$ and $L_y=6$, as shown in Fig. \ref{fig:StepScaleLR6}. 
$F_{4\over 3}(z)$ is correctly recovered in the nonperturbative regime 
as the lattice size is increased 
(when compared with the results of classical Monte Carlo calculations), 
over a wide range of anisotropy $0.45 \leq (a_y/a_x)^6 \leq 0.95$.

\begin{figure}
    \centering
    \includegraphics[width=8.6cm]{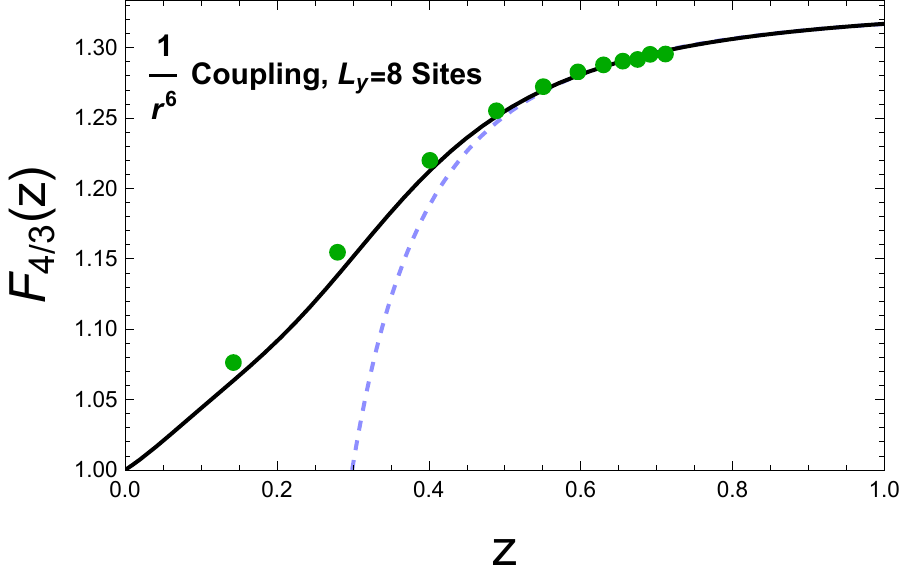}
    \caption{
    $F_{4/3}(z)$  computed by varying $\frac{a_x}{a_y}$  for the $\frac{1}{r^6}$ D-theory Hamiltonian for going from a lattice of size 
    $12\times 8$ sites to $16\times 8$ sites. 
    }
    \label{fig:StepScaleL8}
\end{figure}
To match at scales further into the \ac{UV}, 
lattices with larger $L_y$ must be used. 
However, when $L_y > L_x$ it is possible for dimensional reduction to fail and the $1+1$D O(3) \ac{NLSM} may not be reproduced, as is found for $L_y=8$ where the results overshoot the Monte Carlo and perturbative step-scaling functions, 
as shown in Fig.~\ref{fig:StepScaleYR6}. 
This can be remedied by using lattices with larger $L_x$. 
In Fig. \ref{fig:StepScaleL8}, 
$F_{4/3}(z)$ from $12\times8$ to $16\times8$ lattices with the $\frac{1}{r^6}$ D-theory Hamiltonian is shown, which correctly reproduces the known result
over a larger range than with the $L_y=6$, $\frac{1}{r^6}$ D-theory Hamiltonian. 
This demonstrates how larger correlation lengths may be accessed,
and hence the approach  to the continuum limit.

%%%%%%%%%%%%%%%%%%%%%%%%%%%%%%%%%%%%%%%%%
\section{Quantum Simulations of O(3) \ac{NLSM} using Rydberg Atoms}
Arrays of cold atoms are a promising platform for quantum simulation and as shown above, modest lattice sizes of $6\times6$ and $8\times6$ 
are sufficient to reproduce the \ac{UV} physics of the O(3) \ac{NLSM}, 
and demonstrate asymptotic freedom. This provides an opportunity for a first attempt at performing quantum simulations 
of non-perturbative (\ac{IR}) dynamics of the O(3) \ac{NLSM}. To do so will require the preparation of a low energy state with respect to the Hamiltonian in \cref{eq:DTheoryR6}. The adiabatic spiral~\cite{https://doi.org/10.48550/arxiv.2210.04965}  
can be used to adiabatically prepare the ground state of this Hamiltonian on an array of cold atoms.
\begin{figure}
    \centering
    \includegraphics[width=8.6cm]{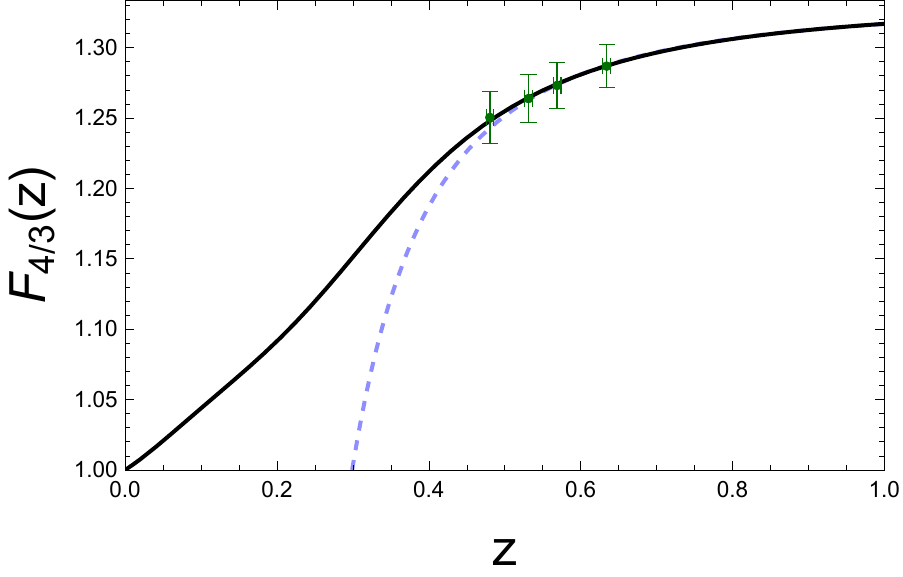}
    \caption{Results for $F_{4/3}(z)$ computed in a TDVP simulation 
    of a rectangular array of $^{87}$Rb atoms assuming 5000 shots are used. }
    \label{fig:StepScaleSpiral}
\end{figure}
To understand the quantum resources required to adiabatically 
prepare states with energy that is  sufficiently 
low to reproduce low-lying physics of the O(3) \ac{NLSM}, 
we performed \ac{TDVP} simulations of the adiabatic spiral using the C++ {\tt ITensor} library~\cite{itensor,itensor-r0.3,  PhysRevLett.107.070601,PhysRevB.94.165116,PhysRevB.102.094315}. 
Details of these calculations can be found in Appendix~\ref{appendix:Rydberg}.
The classical simulations we performed assumed a rectangular array of $^{87}$Rb atoms, 
with $C_6 = 5.42 \times10^6$ MHz $\mu \text{m}^6$, 
with a vertical lattice spacing of $11$ $\mu$m, 
and a selection of horizontal lattice spacings to probe different couplings. 
We assumed a maximum Rabi frequency of $\Omega = 25 \ \text{MHz}$, 
and a maximum coherence time of $4 \mu s$. The initial state of the system with all atoms in their ground state corresponds to a N\`eel state that is degenerate due to a symmetry under reflection of the spins. This degeneracy can be split by evolving with a global detuning term that is turned off during the course of the adiabatic evolution to apply an energy penalty. The initial size of the energy penalty was variationally optimized so that the renormalized coupling of the prepared state matched the vacuum state.
The specific energy penalties and horizontal lattice spacings that we 
used are shown in Tables~\ref{Tab:HardwareParam6} and \ref{Tab:HardwareParam8}. 
Results for the step scaling  obtained from these simulations are shown in Fig.~\ref{fig:StepScaleSpiral}, 
where the uncertainties are derived from 
a sample of 5000 shots in computing the renormalized coupling 
for each lattice configuration. 
\begin{table}[ht]
\centering
  \begin{tabular}{| c | c | c |}
    \hline
    $a_X$ ($\mu$m) & Energy Penalty (MHz) & Final Energy ($\Delta$)\\
    \hline
    12.5 & 0.44 & 2.81 \\
    12.1 & 0.52 & 2.90 \\
    11.8 & 0.56 & 3.43 \\
    11.1 & 0.49 & 4.64 \\
    \hline
  \end{tabular}
  \caption{
  Energy of the ground states prepared using the adiabatic spiral.
  The left column shows the lattice spacing used for the tensor network simulations of a $6\times6$ lattice. The center column shows the energy penalty used to match the vacuum renormalized coupling. The right column shows the energy of the state prepared by the adiabatic spiral in units of the Hamiltonian's energy gap.}
  \label{Tab:HardwareParam6}
\end{table}
\begin{table}[ht]
\centering
  \begin{tabular}{| c | c | c |}
    \hline
    $a_X$ ($\mu$m) & Energy Penalty (MHz) & Final Energy ($\Delta$)\\
    \hline
    12.5 & 0.3  & 4.52  \\
    12.1 & 0.4  & 4.56  \\
    11.8 & 0.46 & 5.43  \\
    11.1 & 0.45 & 7.52  \\
    \hline
  \end{tabular}
  \caption{
    Energy of the ground states prepared using the adiabatic spiral.
The left column shows the lattice spacing used for the tensor network simulations of a $8\times6$ lattice. The center column shows the energy penalty used to match the vacuum renormalized coupling. The right column shows the energy of the state prepared by the adiabatic spiral in units of the Hamiltonian's energy gap.}
  \label{Tab:HardwareParam8}
\end{table}

These simulations show that an ideal cold-atom quantum simulator with only 48 atoms can correctly recover the \ac{UV} physics of the O(3) \ac{NLSM} with sufficient precision. 
To perform this quantum simulation in reality would require 
a rectangular array of $^{87}$Rb atoms with a global driving field 
and a staggered detuning term. 
The parameters used in these simulations are close to those that have 
been implemented in previous cold-atom experiments~\cite{ebadi2021quantum,doi:10.1126/science.abi8794,bernien2017probing,Scholl_2021,Omran_2019}. 
Therefore, it is anticipated that analog quantum simulations of the O(3) \ac{NLSM} should soon be within reach. Due to the similarity to previous cold atom experiments, it is expected that these simulations can be performed with a high degree of fidelity. Scaling to larger systems will require the same pulse sequences applied to larger arrays of atoms. This is not expected to present an issue as larger arrays of Rydberg atoms have been utilized in experiment~\cite{ebadi2021quantum,doi:10.1126/science.abi8794,Scholl_2021} and the techniques used to simulate Heisenberg evolution have been shown to scale to large systems~\cite{https://doi.org/10.48550/arxiv.2207.09438}. Note that while the simulations performed here are for arrays of $^{87}$Rb atoms, similar calculations could be performed using different atomic species, such as Cs~\cite{graham2021demonstration,huft2022simple}.

Reproducing the step scaling curve shows that O(3) \ac{NLSM} physics is actually being reproduced on the quantum simulator and is the first step towards achieving a quantum advantage in the simulation of the O(3) \ac{NLSM}.
Once an approximate vacuum state has been prepared on quantum hardware, 
particle wavepackets can be excited by varying a local detuning or driving term. 
By exciting multiple particles in this manner, scattering in the O(3) \ac{NLSM} can be directly simulated. 
Alternatively, all of this can be also be done at a nonzero $\theta$, by moving the atoms from a rectangular array into a staggered array~\cite{Caspar:2022llo}. 
Using dynamical reconfiguration of atoms, this could even be done dynamically, simulating a quench of the $\theta$ term.
Rapidly turning on $\theta$ would correspond to a rapidly changing axion field~\cite{,PhysRevLett.38.1440,PhysRevLett.40.279} and has been shown to generate a dynamical quantum phase transition in the context of lattice gauge theories~\cite{PhysRevLett.122.050403,https://doi.org/10.48550/arxiv.2210.03089}. 
Both of these calculations involve real-time dynamics that have exponentially scaling computational costs on classical computers, 
and their successful simulation on a quantum computer could represent a true quantum advantage of scientific relevance to high energy physics. 

Note that these problems on the lattice sizes simulated in this section are within the reach of classical computers. Also, a true quantum advantage in simulations of the 1+1D O(3) \ac{NLSM} will need to be performed with a choice of parameters that are outside the reach of perturbation theory. Based on Fig.~\ref{fig:StepScaleLR6}, performing these simulations on a lattice of size $18\times6$ with $\left(a_y/a_x\right)^6=0.45$ is a potential candidate for quantum advantage. A lattice of this size is outside the reach of statevector simulation and lies in the non-perturbative region of the step scaling curve. The DMRG calculations to produce Fig.~\ref{fig:StepScaleLR6} required a bond dimension of 2000 to converge and simulating scattering dynamics or a $\theta$ quench will involve an exponentially growing bond dimension beyond this. Note however, that some tensor networks more suited to 2D such as PEPS may be able to perform this calculation with a lower bond dimension. Regardless, a simulation on this lattice size will be in a regime that is difficult for classical computers and would represent a first chance at seeing a quantum advantage.

%%%%%%%%%%%%%%%%%%%%%%%%%%%%%%
\section{Discussion}

A challenging path lies ahead for the quantum simulation of 
physical systems of importance in high-energy and nuclear physics.
Both Abelian and non-Abelian gauge theories must be mapped efficiently 
onto quantum computers, and
it remains to be determined which of the known frameworks, if any, will evolve toward providing robust predictive capabilities.
For strong interactions, asymptotic freedom has been key in enabling non-perturbative classical calculations with lattice \ac{QCD} of near-static quantities,
and much of the associated technology will translate across to quantum simulations.
In this work, we have studied
a different asymptotically-free field theory.
By developing new methods and performing classical simulations,
we have shown that present-day analog quantum simulators have the potential to perform quantum simulations of 
non-perturbative dynamics within this \ac{QFT} with fully-quantifiable uncertainties.
A definition of the renormalized coupling for the 1+1D O(3) \ac{NLSM} with 
\ac{OBC} was developed to enable the first perturbative matching of 
lattice calculations on quantum simulators to the continuum. 
It is expected that this will enable the use of quantum simulators 
to compute quantities of interest in the continuum limit of the 1+1D O(3) \ac{NLSM}. Additionally, this definition was used to determine the minimal number of qubits 
required for a quantum computer to reproduce continuum physics. 
Remarkably, a cold atom quantum simulator only needs a rectangular array of 48 atoms 
to begin to quantitatively reproduce 
non-perturbative dynamics within the O(3) \ac{NLSM}. 
Cold atoms have been previously used to simulate larger systems and tensor network simulations suggest that existing cold-atom experiments 
should be capable of demonstrating the asymptotic freedom of the O(3) \ac{NLSM}. 
We have also shown that the long-range coupling present in 
cold-atom quantum simulators enables them to make contact with the continuum physics 
of the O(3) \ac{NLSM} with fewer qubits than mappings that 
are restricted to nearest neighbor couplings. 
This is the first concrete example of an ``improved'' Hamiltonian that reduces the qubit count required for a quantum simulation of a lattice field theory to rigorously simulate continuum physics with controlled uncertainties.

While the 1+1D O(3) \ac{NLSM} does not describe any of the fundamental forces in nature, 
it does share a number of qualitative aspects with \ac{QCD} so these simulations will provide valuable insights into how to perform quantum simulations of Standard Model physics.
Our calculations correctly recover the classically-computed step-scaling function, and  demonstrate that the continuum O(3) \ac{NLSM} is being matched, within tolerances,  to  lattices,  and provides new and valuable further steps toward rigorously extracting information about a continuum \ac{QFT} from quantum computers. Once matching has been performed, 
a quantum computer can be used to simulate non-perturbative 
quantities in the theory that are beyond the reach of classical computers, 
including scattering and fragmentation, and $\theta$-quenches.
Further, 
the D-theory mapping studied in this work has the potential to be used to 
simulate the O(3) \ac{NLSM} in $2+1$ dimensions by making use of 
3D cold-atom arrays which have recently been experimentally 
demonstrated~\cite{Barredo_2020}.

%%%%%%%%%%%
\section*{Acknowledgments}

We would like to thank Pavel Lougovski and Peter Komar from the AWS Braket team for useful discussions. The views expressed are those of the authors and do not reflect the official policy or position of AWS. We would also like to thank Henry Froland and Nikita Zemlevskiy for helpful discussions.
This work was supported
in part by U.S. Department of Energy, Office of Science, Office of Nuclear Physics, Inqubator for Quantum Simulation (IQuS) under Award Number DOE (NP) Award DE-SC0020970,
in part by the DOE QuantISED program through the theory  consortium ``Intersections of QIS and Theoretical Particle Physics'' at Fermilab with Fermilab Subcontract No. 666484,
and in part by Institute for Nuclear Theory with US Department of Energy Grant DE-FG02-00ER41132.
This work was enabled, in part, by
the use of advanced computational, storage and networking infrastructure provided by the Hyak supercomputer system at the University of Washington 
\footnote{\url{https://itconnect.uw.edu/research/hpc}}.
This work was also supported, in part, through the Department of Physics 
\footnote{\url{https://phys.washington.edu}}
and the College of Arts and Sciences 
\footnote{\url{https://www.artsci.washington.edu}}
at the University of Washington. This work also made use of AWS EC-2 compute instances through the generous support of AWS Programs for Research and Education.

%%%%%%%%%%%%%%%%%%%%%%%%%%%
\clearpage
\appendix
\section{Rydberg Atom Simulation}
\label{appendix:Rydberg}
The Hamiltonian describing the evolution of a 
rectangular array of Rydberg atoms is
\begin{align}
    \hat{H}^\text{Ryd.}(t) & = \nonumber  \\
    & \sum_{x_1,y_1,x_2,y_2} \frac{C_6\ \  \hat{n}_{x_1,y_1} \hat{n}_{x_2,y_2} }{\left( a_x^2(x_1-x_2)^2 + a_y^2(y_1-y_2)^2 \right)^3} \nonumber \\
    & + \sum_{x,y} \Delta_{x,y}(t) \hat{n}_{x,y} 
    + \sum_{x,y} \frac{\Omega_{x,y}(t)}{2} \hat{X}_{x,y}  \ \ \ ,
    \label{eq:RydbergArray}
\end{align}
where $\hat{n}_{x,y}$ is the Rydberg occupation number, 
$\Delta_{x,y}(t)$ is a position dependent detuning term, 
$\Omega_{x,y}(t)$ is a position dependent driving term, 
$a_x$ is the horizontal lattice spacing and 
$a_y$ is the vertical lattice spacing. 
As presented in the main text, 
the Rydberg number operator can be identified with a staggered spin operator, i.e., $\hat{n}_{x,y} = \frac{1}{2} + (-1)^{x+y}\hat{S}^z_{x,y}$, 
such that the state with all atoms in their ground state corresponds 
to a N\'{e}el state. 
With this identification, the adiabatic spiral introduced in Ref. \cite{https://doi.org/10.48550/arxiv.2210.04965} 
can be used to prepare a 
low energy state of the Hamiltonian in Eq.~(\ref{eq:DTheoryR6}),
by using
\begin{align}
   \Delta_{x,y}(t) & = (-1)^{x+y}\Omega_D + h_P \left(1 - \frac{t}{T}\right) \nonumber \\
   & + \frac{1}{2}\sum_{(x_2,y_2)\neq (x,y) } \frac{C_6}{\left( a_x^2(x-x_2)^2 + a_y^2(y-y_2)^2 \right)^3} 
   \ ,
   \nonumber\\
   \Omega_{x,y}(t) & = \sqrt{2}\  \Omega_D \left(\frac{t}{T} + \frac{1}{\pi} \sin\left(\pi \frac{t}{T}\right)\right) \ \ \ ,
   \label{eq:SpiralFields}
\end{align}
where $h_P$ is an initial energy penalty, 
$\Omega_D$ specifies the final strength of the driving field, 
and $T$ is the total time used for the adiabatic state preparation. 
For our calculations, we have used 
$\Omega_D = \frac{1}{\sqrt{2}} 25 \ \text{MHz}$, $T=3.83 \mu s$, 
and $h_P$ is presented in Tables~\ref{Tab:HardwareParam6} 
and \ref{Tab:HardwareParam8}. 
Performing a measurement on a Rydberg atom simulator requires the 
drive field to be turned off,
which we simulated by quenching
$\Omega_{x,y}(t)$  to zero over 
a time interval of $0.1 \ \mu \text{s}$. 
We assumed that a combined time of $0.07 \ \mu \text{s}$ 
was required to turn the detuning on and off.

The adiabatic spiral described here was simulated with tensor networks. This was done with the C++ {\tt iTensor} library with {\tt OpenBLAS} 
as the backend to parallelize the linear algebra operations~\cite{itensor}. 
The state of the system was represented with a matrix product state (MPS) tensor network that wound through the 2D lattice. Time evolution was performed by discretizing $\hat{H}^\text{Ryd.}(t)$ into 200 time independent steps and evolving with 1-site TDVP~\cite{PhysRevLett.107.070601,PhysRevB.94.165116}. 
Before each step, the bond dimension was increased using the global Krylov method~\cite{PhysRevB.102.094315}, with a maximum allowed bond dimension of 550.

\bibliography{refs,bibi,biblioKRS}

\end{document}